\documentclass[
 reprint,
 amsmath,amssymb,
 aps,
 prl,
 floatfix,
]{revtex4-2}

\usepackage[T1]{fontenc}
\usepackage[utf8]{inputenc}
\usepackage{graphicx}
\usepackage{bm}
\usepackage{xcolor}
\usepackage{booktabs}
\usepackage{mathtools}
\usepackage{soul}
\usepackage[colorlinks=true,linkcolor=blue,citecolor=blue,urlcolor=blue]{hyperref}

\newcommand{\ii}{\mathrm{i}}

\begin{document}

\title{Anisotropic optical chirality and Lipkin's zilch tensor}

\author{Ilia Smagin}
\email{Ilia.Smagin@skoltech.ru}

\author{Sergey Dyakov}
\email{S.Dyakov@skoltech.ru}

\affiliation{Skolkovo Institute of Science and Technology, Moscow, Russia}

\date{\today}

\begin{abstract}
Optical chirality density is widely used as a scalar measure for describing the chiral properties of electromagnetic fields and their interaction with isotropic chiral media. However, in anisotropic optically-active media, a single scalar quantity is generally insufficient to capture the full complexity of chiral field-matter coupling. In this work, we go beyond the conventional optical chirality density and introduce a tensor of electromagnetic chirality based on the Lipkin formalism. This tensor provides a richer and more physically transparent description of chiral electromagnetic fields, particularly in the context of their interaction with general magneto-electric media. We discuss the physical meaning of the individual tensor components and provide an expression for the excitation rate of small anisotropically chiral molecules in the presence anisotropically chiral fields.

\end{abstract}

\maketitle


Chiral optical response is one of the clearest manifestations of mirror-symmetry breaking in electrodynamics. Its description requires field quantities with transformation properties different from those of energy, momentum, or angular momentum. At the electric-dipole--magnetic-dipole level, the electromagnetic quantity responsible for isotropic enantiomer-sensitive absorption must be even under time reversal and odd under spatial inversion. The standard local quantity with these properties is an optical chirality density introduced by Tang and Cohen~\cite{Tang2010}. 
Along with the associated chirality flux, these two quantities satisfy the conservation law
which makes optical chirality a natural scalar descriptor of electromagnetic handedness.

The scalar character of optical chirality density however, is also its limitation. Namely, this quantity is the appropriate field-side object only when the electric--magnetic response of matter is isotropic. In oriented molecules, crystals, metamaterials, or structured photonic crystals, optical activity is generally described not by a pseudoscalar but by a pseudotensor. The electromagnetic quantity coupled to such a material response cannot, in general, be exhausted by a single scalar density. It must contain directional information while retaining the same time-reversal and parity properties required for chiral response.

It appears that this tensorial structure is already present in existing formalism of Lipkin's zilch tensor~\cite{Lipkin1964}. Lipkin introduced a third-rank electromagnetic pseudotensor \(Z^{\mu\nu\sigma}\) with $\mu,\nu,\sigma=0,1,2,3$ whose components obey local conservation laws in free space. Historically, the remaining zilch components were regarded mainly as mathematically conserved quantities with obscure physical content. This ambiguity is reflected already in the name ``zilch''. Later works clarified the symmetry origin and conservation properties of these quantities~\cite{kibble1965conservation,Morgan1964,Candlin,Simulik1989_1,Simlulik1989_2,Philbin2013,aghapour2020zilch,letsios2022continuity}, while the Tang--Cohen result supplied a physical interpretation for the scalar component $Z^{000}$ as the optical chirality density. 

From the classical definition~\cite{Lipkin1964}, the Lipkin densities $Z^{ab0}$ are expessed as
\begin{align}
\begin{split}
&Z^{ab0} = \delta_{ab}\Bigg[\bm E\cdot rot\bm E + \bm H\cdot rot\bm H \Bigg]- \\
&- E_a (rot \bm E)_b - H_a (rot \bm H)_b - E_b (rot \bm E)_a - H_b (rot \bm H)_a.
\end{split}
\end{align}
For monochromatic fields with the time dependence $\exp(-\mathrm{i}\omega t)$, the diagonal components of time-averaged Lipkin densities $Z^{ab0}$ ($a,b=1,2,3$) are
\begin{equation}
\begin{aligned} \overline{Z^{xx0}} &= \omega\,\operatorname{Im} \left( -E_xH_x^{*}+E_yH_y^{*}+E_zH_z^{*} \right), \\
\overline{Z^{yy0}} &= \omega\,\operatorname{Im} \left( E_xH_x^{*}-E_yH_y^{*}+E_zH_z^{*} \right), \\
\overline{Z^{zz0}} &= \omega\,\operatorname{Im} \left( E_xH_x^{*}+E_yH_y^{*}-E_zH_z^{*} \right),
\end{aligned}\label{eq:Zab0_diagonal}
\end{equation}
while off-diagonal components are
\begin{equation}
\begin{aligned} \overline{Z^{xy0}} = \overline{Z^{yx0}} &= -\omega\,\operatorname{Im} \left( E_xH_y^{*}+E_yH_x^{*} \right), \\
\overline{Z^{xz0}} =\overline{Z^{zx0}} &= -\omega\,\operatorname{Im} \left( E_xH_z^{*}+E_zH_x^{*} \right), \\
\overline{Z^{yz0}} = \overline{Z^{zy0}} &= -\omega\,\operatorname{Im} \left( E_yH_z^{*}+E_zH_y^{*} \right).
\end{aligned} \label{eq:Zab0_offdiagonal}
\end{equation}
The conventional optical chirality density is recovered up to a factor of $\frac12$ as the trace of $\overline{Z^{ab0}}$
\begin{equation}
\overline{Z^{000}} = \sum_{a=1}^{3}\overline{Z^{aa0}} = \omega\,\operatorname{Im} \left(\bm E\cdot\bm H^{*}\right). \label{eq:Z000_trace}
\end{equation}
Beyond the optical chirality density, Yang and Cohen related nine components $\overline{Z^{ab0}}$, $a,b=1,2,3$ to the bilinear field factors governing electric-dipole--magnetic-dipole transitions in oriented molecules~\cite{yangcohen2011}. However, their expression for the excitation rate retained the full mixed electric--magnetic response tensor and therefore did not separate the symmetric Lipkin-related contribution from the antisymmetric, nonchiral one.

Building on this result, we treat separately the trace and the traceless symmetric and antisymmetric parts of the mixed electric--magnetic pseudotensor. We demonstrate while the trace part couples to classical optical chirality density, the traceless symmetric part is probed by the five Lipkin densities, which form part of the Lipkin tensor. On the other hand, for the antisymmetric part of the optical-activity pseudotensor, the coupling term is proportional to the reactive part of the local Poynting vector. 
This formulation includes both diagonal and off-diagonal contributions and clearly separates the conventional isotropic optical-chirality response from its direction-dependent tensorial extension. The remaining Lipkin densities therefore characterize magnetoelectric responses that cannot be distinguished by a scalar field measure.



We illustrate this result by calculating the absorption spectra of a cavity supporting fields with non-zero off-diagonal Lipkin densities $Z^{ab0}$. We demonstrate that when the cavity is filled with a magneto-electric medium, the optical response is odd under $G_{ab}\to -G_{ab}$ if only some of the optical-activity tensor components $G^{ab}$ match certain Lipkin densities $Z^{ab0}$.


Following the Tang--Cohen approach, we consider an optically-active molecule described at the electric-dipole--magnetic-dipole level by
\begin{align}
 p_a &= \alpha_{ab}E_b-\ii G_{ab}H_b,
&
 m_a &= \chi_{ab}H_b+\ii G_{ba}E_b .
\label{eq:dipoles}
\end{align}
Here $\alpha_{ab}$ and $\chi_{ab}$ are the electric polarizability and magnetic susceptibility tensors, while $G_{ab}$ is the optical-activity pseudotensor~\cite{Fedorov,Barron}. In what follows, our goal is to analyze the optical properties of the system for two opposite pseudotensors $G_{ab}$ at a given field configuration. Please note that the optical-activity pseudotensor $G_{ab}$ with non-zero components may correspond not only to a chiral medium but also pseudochiral medium, omega-medium or general-bianisotropic medium. For a chiral medium, the opposite-sign $G_{ab}$ tensors correspond to two enantiomers. For a pseudochiral medium, the opposite $G_{ab}$ tensors describe the same medium but with different orientations.

Substituting Eq.~\eqref{eq:dipoles} into the formula for the time-averaged excitation rate of the molecule
\begin{equation}\label{eq:Aiso}
    A^\pm = \langle \mathbf{E}\cdot\dot{\mathbf{p}} + \mathbf{H}\cdot\dot{\mathbf{m}}\rangle = \frac{\omega}{2}Im(\mathbf{E}^*\cdot\mathbf{p} + \mathbf{H}^*\cdot\mathbf{m})
\end{equation}
and separating the right-hand side into terms independent of $G_{ab}$ and terms linear in $G_{ab}$, one can obtain an expression for the excitation rates at opposite $G_{ab}$ (see Ref.~[\citenum{smagin2026beyond}] for details). Next, following the procedure used in classification of bi-anisotropic media \cite{tretyakov1998magnetoelectric}, we decompose the general pseudotensor $G_{ab}$ into trace part and traceless symmetric and antisymmetric parts
\begin{align}\label{eq:G_Tretyakov_decomposition}
    G_{ab}
    =
    G\delta_{ab}
    +
    G_{(ab)}
    +
    G_{[ab]},
\end{align}
where
\begin{align}
\begin{split}
    &G
    =
    \frac13 G_{aa},
    \\
    &G_{(ab)}
    =
    \frac12
    \left(
        G_{ab}
        +
        G_{ba}
    \right)
    -
    G\delta_{ab},
    \\
    &G_{[ab]}
    =
    \frac12
    \left(
        G_{ab}
        -
        G_{ba}
    \right)
    =
    \epsilon_{abc}\Omega_c,
\end{split}
\end{align}
which gives the central result
\begin{align}\label{eq:central_result}
    A^\pm
    =
    A_0
    \mp
    G''
    \overline{Z^{000}}
    \pm
    \frac12
    G_{(ab)}''
    \overline{Z^{(ab)0}}
    \pm
    \omega
    \bm\Omega''
    \cdot
    \operatorname{Im}
    \left(
        \mathbf E^*
        \times
        \mathbf H
    \right)
\end{align}
Here the term $A_0$ collects the ordinary electric and magnetic excitation terms, $G = \frac13\mathrm{tr}\{G_{ab}\}$ is the pseudoscalar magneto-electric parameter,  $G_{(ab)}$ denotes symmetric traceless part of the $G_{ab}$ pseudotensor, and $G_{[ab]}$ is the antisymmetric traceless part. Note that the latter can always be equivalently represented via the polar vector $\bm\Omega = \frac{1}{2}\epsilon_{abc}G_{ab} =  [G_{[23]}, -G_{[13]}, G_{[12]}]$.

In this terminology, the first term in \eqref{eq:G_Tretyakov_decomposition} describes isotropic chiral medium (Pasteur medium), the second term corresponds to anisotropic chiral or pseudochiral medium, while the last term represents so-called omega medium. All terms together describe general bi-anisotropic medium.

\begin{figure*}[t]
    \centering
    \includegraphics[width=1\linewidth]{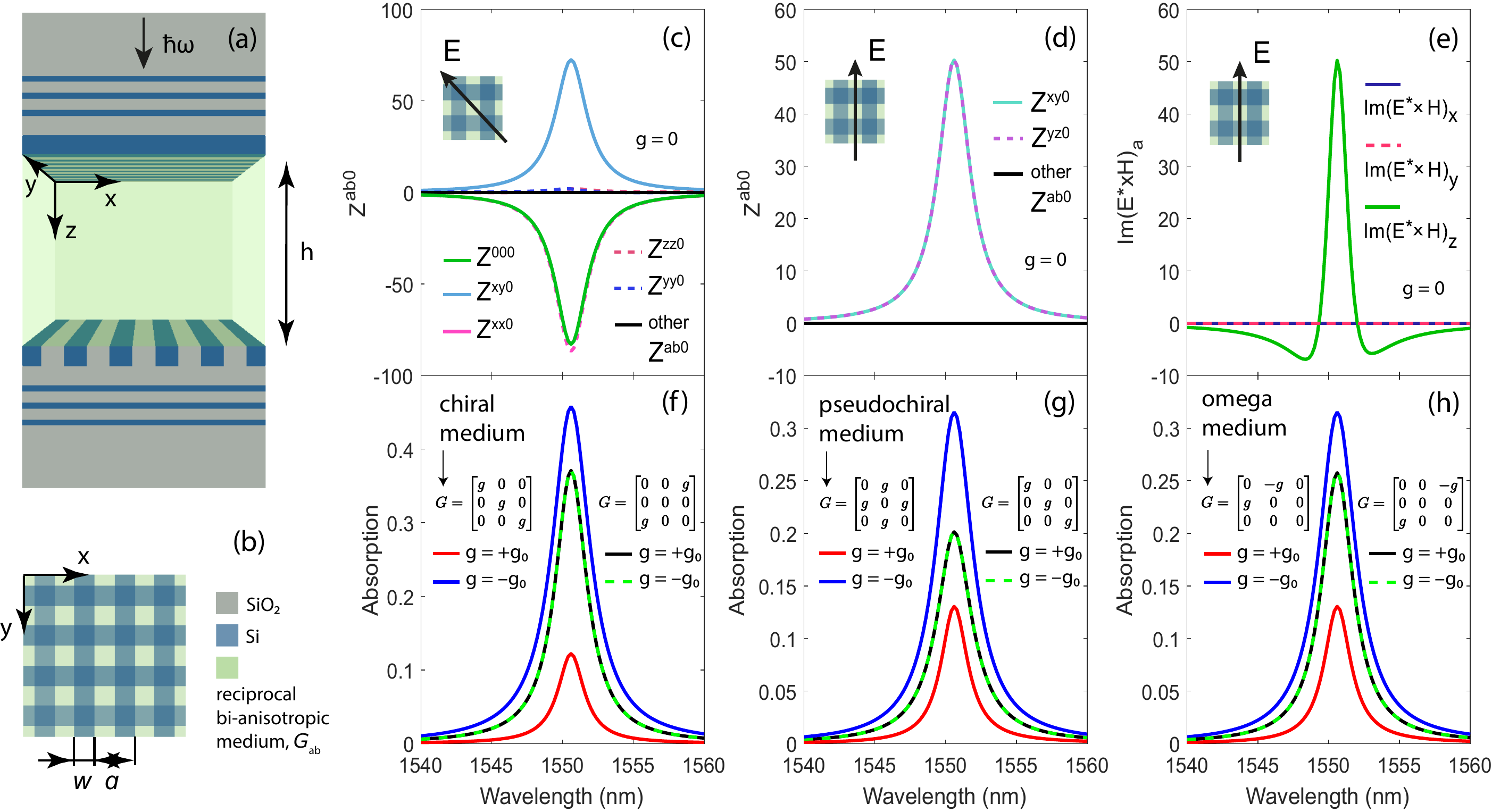}
    \caption{(a) Schematic of the resonator supporting electromagnetic fields with non-zero Lipkin densities $\overline{Z^{ab0}}$ and reactive part of the Poynting vector. (b) Top view of crossed one-dimensional gratings. (c,d) Spectral dependencies of various Lipkin densities $\overline{Z^{ab0}}$ and (e) components of reactive part of Poynting vector, calculated for normally incident light; (f,g,h) absorption spectra of a resonator containing different reciprocal bi-anisotropic materials. Panels (c,d,f,g) correspond to polarization along diagonal of the square unit cell of the resonator, while panels (e,h) correspond to polarization along the $y$-axis. Colors in (a) and (b) denote different materials. Gap size $h = 200$~nm. Thicknesses of layers in mirrors are the following: 636~nm (grating), 112~nm, 114~nm, 276~nm, 114~nm, 276~nm, and 114~nm. Parameters of bi-anisotropic material are the following: $\varepsilon = 2+0.01i$, $G^\mathrm{xy}_\mathrm{c2v}= \left(\begin{smallmatrix} 0&g&0 \\ g&0&0 \\ 0&0&0 \end{smallmatrix}\right)$, $G^\mathrm{xz}_\mathrm{c2v}= \left(\begin{smallmatrix} 0&0&g \\ 0&0&0 \\ g&0&0 \end{smallmatrix}\right)$; $G_\mathrm{O}= \left(\begin{smallmatrix} g&0&0 \\ 0&g&0 \\ 0&0&g \end{smallmatrix}\right)$, $g = 0.003i$. }
    \label{absorption_dif}
\end{figure*}

\begin{table}[b]
\centering
\renewcommand{\arraystretch}{1.25}
\begin{tabular}{p{0.2\columnwidth}|p{0.4\columnwidth}|p{0.34\columnwidth}}
\hline
Symmetry group & Allowed nonzero components of $G''_{(ab)}$&{Class of medium}\\
\hline
$C_1$ 
& all components {$G''_{(ab)}$} & {general bianisotropic}\\
\hline
$C_2$ 
& {$G''_{xx},G''_{yy},G''_{zz},G''_{xy}$}& {anisotropic chiral}\\
\hline
$C_s$ 
& {$G''_{xz},G''_{yz}$}& {pseudochiral}
\\
\hline
$C_{2v}$ 
& {$G''_{xy}$} & {pseudochiral}
\\
\hline
$D_2$ 
&{$G''_{xx},G''_{yy},G''_{zz}$} & {anisotropic chiral}
\\
\hline
{$C_n$ ($n\geq3$)}
& {$G''_{xx}=G''_{yy},G''_{zz}$} & {anisotropic chiral}
\\
\hline
{$D_n$ ($n\geq3$)}
& {$G''_{xx}=G''_{yy},G''_{zz}$} & {anisotropic chiral}
\\
\hline
$S_4$ 
& {$G''_{xx}=-G''_{yy},G''_{xy}$} & {anisotropic chiral}
\\
\hline
$D_{2d}$ 
& {$G''_{xy}$} & {pseudochiral}
\\
\hline
$T,O$ 
& {$G''_{xx}=G''_{yy}=G''_{zz}$} & {isotropic chiral}
\\
\hline
\end{tabular}
\caption{Relation between selected point groups~\cite{LandauVol8}, their symmetry operations, and the allowed nonzero components of the symmetric optical-activity pseudotensor $G''_{(ab)}$. The principal symmetry axis is chosen along $z$; for $C_s$, the mirror plane is the $xy$ plane. The rows $C_n$ and $D_n$ with $n\geq3$ include noncrystallographic molecular groups such as $C_5$ and $D_5$.}
\label{tab:Gsym}
\end{table}

Equation~\eqref{eq:central_result} is the generalization of the Tang--Cohen relation to the anisotropic case. In the isotropic case, the enantiomer-sensitive part of the excitation rate is proportional to the product of a scalar material parameter and the scalar optical chirality density. 
The isotropic limit is recovered when $G''_{ab}=G''\delta_{ab}$. For such $G''_{ab}$, only the trace contribution remains in the asymmetric magneto-optical response, and Eq.~\eqref{eq:central_result} reduces to the familiar scalar coupling proportional to the conventional optical chirality density $\overline{Z^{000}} = \omega \mathrm{Im}(E\cdot H^*)$. 

Therefore, equation~\eqref{eq:central_result} gives a simple practical interpretation of the remaining Lipkin zilch components. An isotropic optically active molecule is sensitive only to the trace of the tensor $\overline{Z^{ab0}}$, which gives the usual optical chirality density. By contrast, an oriented anisotropic molecule, an optically active crystal, or a bianisotropic meta-atom can distinguish different Lipkin densities. Such a system may therefore show a pronounced difference in optical responses for opposite $G_{ab}$ even when the scalar optical chirality has vanished. In this sense, the remaining Lipkin zilch components acquire a clear physical meaning: now they are not formally conserved quantities but are the optical density components coupled to anisotropic optical activity pseudotensor.


An important remark has to be made here. As mentioned, although Eq.~\eqref{eq:central_result} was derived by extending the Tang--Cohen treatment of enantiomer-sensitive excitation to anisotropic molecules, its range of validity is not limited to purely chiral matter. Indeed, as Table~\ref{tab:Gsym} shows, nonzero components of the optical-activity pseudotensor $G_{ab}$ may also be symmetry-allowed even for pseudochiral molecules possessing improper symmetry operations. In such cases, the contribution that is odd under the formal transformation ($G_{ab}\to -G_{ab}$) still describes a genuine mixed electric--magnetic response and is still governed by the corresponding Lipkin densities. This, however, should not be interpreted as enantiomer-sensitive response. Thus, Eq.~\eqref{eq:central_result} identifies the Lipkin densities as field-side descriptors of the more general anisotropic magnetoelectric coupling, with pseudoscalar optical activity parameter $G$ emerging as an important special case.


We now illustrate Eq.~\eqref{eq:central_result} with a resonant photonic structure designed to enhance selected Lipkin densities $\overline{Z^{ab0}}$ and components of the reactive part of the Poynting vector $\mathrm{Im}(\mathbf E^*\times \mathbf H)$. The system is a Bragg microresonator formed by a pair of orthogonal one-dimensional gratings on a multilayered mirror; the cavity region of this microresonator can be filled with magneto-electric material as shown in Fig.~\ref{absorption_dif}(a). Such a resonator without magneto-electric material has a doublet of chiral modes in the $\Gamma$-point of the photonic crystal lattice \cite{dyakov2025strong}. By choosing appropriate polarization of the normally incident wave (along the $y$-axis or along the diagonal of the square unit cell), one can excite the field in the cavity region with different $\overline{Z^{ab0}}$ and $\mathrm{Im}(\mathbf E^*\times \mathbf H)_a$. The purpose of this example is to show that Lipkin densities \and components of reactive Poynting vector can be selectively coupled to corresponding components of the material pseudotensor $G''_{ab}$. 

Figure~\ref{absorption_dif}(c) shows the spectrum of $\overline{Z^{ab0}}$ for incident polarization along the diagonal of the unit cell. In such configuration, in the spectral region near the cavity resonance, the dominant non-zero densities are $\overline{Z^{000}}$, $\overline{Z^{zz0}}$ and $\overline{Z^{xy0}}$, whereas the remaining density components are strongly suppressed. This is a classical example of chiral field and, according to Eq.~\eqref{eq:central_result}, it should couple efficiently to the material, whose optical-activity pseudotensor when oriented properly contains non-zero components $G''_{zz}$ and/or $G''_{xy}=G''_{yx}$. This prediction is confirmed by the calculated absorption spectra shown in Fig.~\ref{absorption_dif}(f). As one can see from Fig.~\ref{absorption_dif}(f), introducing an isotropic chiral O-symmetric material with $G=\pm g_0\delta_{aa}$ yields in absorption resonant peaks with amplitudes, different for opposite material enantiomers. By contrast, the optical-activity pseudotensor with only non-zero components $G_{13}=G_{31}$ (C$_{2v}$ symmetry group) does not couple to the excited field with diagonal polarization, because the corresponding third term in Eq.~\eqref{eq:central_result} vanishes. 

A complementary situation is shown in Fig.~\ref{absorption_dif}(d), where the incident polarization is directed along the $y$-axis. In this case, the excited field produces sizable off-diagonal densities $\overline{Z^{xy0}}$ and  $\overline{Z^{yz0}}$, whereas the scalar density $\overline{Z^{000}}$ (equal to the trace of $\overline{Z^{ab0}}$), as well as the rest densities, are negligible. In contract to the previous case, such a field is no longer chiral and thus interacts equally with opposite enantiomers of the isotropic chiral material [see black and green lines in Fig.~\ref{absorption_dif}(g)]. At the same time, according to Eq.~\eqref{eq:central_result}, this field should couple efficiently to the optical-activity pseudotensor containing the corresponding off-diagonal components $G''_{xy}=G''_{yx}$ and/or $G''_{yz}=G''_{zy}$. Indeed, introducing such a material into the cavity region, gives the pronounced difference in the resonance absorption amplitudes for opposite material orientations [red and blue lines in Fig.~\ref{absorption_dif}(g)]. Therefore, in this particular example, the absorption spectrum is sensitive only to the selected off-diagonal components of the material optical activity pseudotensor rather than to the entire pseudotensor or to the scalar optical activity parameter $G$ used in \cite{Tang2010}.

A similar strategy can be applied to couple the resonator field to an omega-medium, whose magnetoelectric response is described by a pseudotensor $G''_{ab}$ with off-diagonal components of opposite sign, $G''_{ab} = -G''_{ba}$. According to Eq.~\eqref{eq:central_result}, such a medium does not couple to Lipkin density componets, but it can interact efficiently with fields that possess an enhanced non-zero reactive part of the Poynting vector, $\mathrm{Im}(\mathbf E^*\times \mathbf H)_a$. To achieve this, we again use the incident polarization along the $y$-axis, which, as shown in Fig.~\ref{absorption_dif}(e), excites a field in the cavity with a strongly enhanced component $\mathrm{Im}(\mathbf E^*\times \mathbf H)_3$. When an omega-medium with non-zero components $G''_{12} = -G''_{21}$ is introduced into the cavity, the fourth term in Eq.~\eqref{eq:central_result} becomes active, leading to a pronounced difference in the absorption spectra for opposite orientation of the omega-medium, as confirmed by the calculated spectra shown in Fig.~\ref{absorption_dif}(h). In contrast, an omega-medium with components $G''_{13} = -G''_{31}$ does not couple to this field, because the reactive Poynting vector component along the $z$-axis is zero for this excitation, and thus no differential absorption is observed. This example demonstrates that, just as for chiral media, the selective enhancement of specific reactive Poynting vector components enables the resonator to discriminate between different orientations of the omega-medium pseudotensor.

These results demonstrate the practical content of Eq.~\eqref{eq:central_result}. The dipolar response to external field is controlled not by the existence of a chiral field and a non-zero material optical-activity pseudoscalar $G$ but by the \textit{overlap} between the elements of the material optical-activity pseudotensor $G_{ab}$ and the matching components of the Lipkin field tensor and/or reactive part of the Poynting vector. As a result, the anisotropic optical activity may remain optically silent if the required component is not present in the field. The usual Tang--Cohen scalar coupling is therefore recovered as the isotropic trace projection of a more general tensorial relation.


In conclusion, we have established a clear physical interpretation for the components of the zeroth slice of the Lipkin zilch tensor: they represent the electromagnetic field densities that directly couple to the symmetric part of a general optical-activity pseudotensor. We also have demonstrated that the ansisymmetric part of this tensor is coupled to the reactive part of the Poynting vector. This identification generalizes the conventional Tang--Cohen optical chirality density, which is recovered as the isotropic limit. Our results thus provide a rigorous framework for analyzing chiral, pseudochiral and omega-like light--matter interactions beyond isotropic media, with implications for structured light fields and anisotropic chiral materials.

\textit{Acknowledgments}---This work was supported by the RSF, project 25-12-00454. The authors thank N.~Gippius, I.~Fradkin and V.~Fedotov for valuable discussions.

\bibliography{ref}

\end{document}